\documentclass [prb,superscriptaddress,floatfix,showpacs] {revtex4}

\usepackage {amsmath,amssymb,epsfig,color}

\begin{document}

\title
{Crystal-field splitting for low symmetry systems in ab initio calculations}

\author{S.V.~Streltsov}
\affiliation{Ural State Technical University, Mira St. 19, 620002 Ekaterinburg, Russia}
\affiliation{Institute of Metal Physics, S.Kovalevskoy St. 18, 620219 Ekaterinburg GSP-170, Russia}
\email{streltsov@optics.imp.uran.ru}

\author{A.S.~Mylnikova}
\affiliation{Ural State Technical University, Mira St. 19, 620002 Ekaterinburg, Russia}
\affiliation{Institute of Metal Physics, S.Kovalevskoy St. 18, 620219 Ekaterinburg GSP-170, Russia}
\author{A.O.~Shorikov}
\affiliation{Institute of Metal Physics, S.Kovalevskoy St. 18, 620219 Ekaterinburg GSP-170, Russia}
\author{Z.V.~Pchelkina}
\affiliation{Institute of Metal Physics, S.Kovalevskoy St. 18, 620219 Ekaterinburg GSP-170, Russia}
\author{D.I.~Khomskii}
\affiliation{II. Physikalisches Institut, Universit$\ddot a$t zu K$\ddot o$ln,
Z$\ddot u$lpicher Stra$\ss$e 77, D-50937 K$\ddot o$ln, Germany}
\affiliation{Groningen University, Nijenborgh 4, 9747 AG Groningen, The
Netherlands}
\author{V.I.~Anisimov}
\affiliation{Institute of Metal Physics, S.Kovalevskoy St. 18, 620219 Ekaterinburg GSP-170, Russia}

\date{\today}

\begin{abstract}
In the framework of the LDA+U approximation we propose the direct way 
of calculation of crystal-field excitation energy and
apply it to La and Y titanates. The method developed can be 
useful for comparison with the results of spectroscopic measurements 
because it takes into account fast relaxations of electronic system. 
For titanates these relaxation processes reduce the value of 
crystal-field splitting by $\sim30\%$ as compared with 
the difference of LDA one electron energies. However, the 
crystal-field excitation energy in these systems is still large enough 
to make an orbital liquid formation rather unlikely
and experimentally observed isotropic magnetism remains
unexplained.
\end{abstract}

\maketitle

\section{Introduction}
\label{intro}
The magnitude of crystal-field splitting (CFS) is known to be very
important characteristic of transition metal compounds, necessary
for understanding different physical phenomena such as magnetism
and metal-insulator transitions.\cite{Goodenough,Ballhausen} 
It is
often used for detailed fitting of spectroscopic data and in
various model calculations. An important question is: how can one
calculate CFS in a reliable way?

One can obtain character of CFS using the group theory analysis.
It provides qualitative description of ionic levels structure in 
the presence 
of crystal-field of given symmetry and can even give some ratios between 
splittings using Wigner-Eckart theorem.\cite{Elliott}
However, for quantitative estimates the detailed structure of ionic
potentials has to be taken into account.

Generally, there are different contributions to the total value of 
CFS to be considered. A point-charge
contribution can be calculated ``by hand''  using for
instance direct Madelung potential summation (see e.g.~\cite{Cwik-03}).
However, this method has some drawbacks and neglects the 
overlap of the electron clouds of neighboring ions, which can lead to 
several important consequences.\cite{Goodenough}

Another important contribution comes from the covalency. It can also be
estimated using some model calculations (for instance taking Slater-Koster
parameters), but without precise knowledge of the band structure one 
does not actually know the parameters of the models accurately enough.
 In order to take into consideration all these contributions
to the total value of CFS in a reliable way one should use {\it
ab initio} methods based on density functional theory (DFT)
calculations. Nevertheless, even in this case there are several ways to
define and calculate CFS. Each of them has its own virtues, but
also difficulties.

The first possibility is a direct calculation of the center of
gravity of corresponding bands.  It works well when 
splitting is large enough. Thus it is most useful and rather
accurate for $t_{2g}-e_g$ splitting (typically in 3d-oxides CFS 
between $t_{2g}$ and $e_g$ states is $\sim$2-2.5~eV). For example the behavior of 
spin-state transition temperature in cobaltites, which depends on 
this splitting has been recently estimated quite accurately.\cite{Nekrasov-03} 
The problems might arise in computation of smaller values such as the
splitting inside $t_{2g}$ or $e_g$ shells. 
The general problem is
that CFS is a well-defined characteristic only for isolated ions
which have energy levels but not bands. The bigger the corresponding 
band width (in comparison with CFS) the more questionable it becomes to treat
the difference of centers of gravity as an estimate of CFS
value.

The second way to calculate CFS is the downfolding or projection
procedure which could give a few orbital on-site Hamiltonian in the
minimal basis set of functions  $\psi_1$, $\psi_2$, 
..., $\psi_N$.\cite{Andersen-00,Anisimov-04}
In Sec.~IV the method of obtaining such few orbital Hamiltonian 
from a full orbital one is briefly described.

Having this Hamiltonian one
can diagonalize it to obtain its eigenvalues $\epsilon_1$, $\epsilon_2$,
..., $\epsilon_N$ and eigenvectors
$\Psi_1$, $\Psi_2$, ..., $\Psi_N$ which are in general linear
combinations of the initial wave functions:
\begin{equation}
\label{Psi}
|\Psi_j> = \sum^N_{i} a_{ij} |\psi_i>.
\end{equation}
The eigenvalues can be considered as the energies of the orbitals. 
The differences between these eigenvalues define in this case CFS.

In spite of great generality of this method, it has some
disadvantages. There is an ambiguity in the definition of unitary
transformation matrix $U^{({\bf k})}_{ji}$ for Wannier function
construction (see Sec. IV) 
and in the choice of the basis set wave functions 
$\psi_1$, $\psi_2$, ..., $\psi_N$ in any downfolding or projection
procedure. Using different basis sets few orbital non-interacting 
Hamiltonian can be constructed in different ways. All of 
them are equally good if the resulting Hamiltonian gives the same 
bands as the LDA full orbital one. It is essentially important 
for low symmetry systems, where it is not clear what 
kind of linear 
combinations of $d-$wave functions and in which local coordinate 
system (LCS) should be taken as a basis set for few orbital Hamiltonian 
construction.

In addition there is one more serious and general problem.  
LDA calculations are widely known to give good description
of the {\it ground state} characteristics. However, it often
fails to describe the excited states. LDA eigenvalues
should not be treated as the real excitation energies. 
This general shortcoming concerns all methods of 
calculating CFS which use LDA eigenvalues.

On the other hand, care should be taken in what we actually mean
by CFS.  In direct study of CF excitations e.g. by optics
one should take into account possible relaxation of both the
electronic (fast process) and lattice (relatively slow relaxation)
subsystems.  Fast electronic relaxation definitely has to be taken 
into consideration,  whereas the lattice
relaxation can be often treated separately (direct optical
absorptions usually occur at a frozen lattice,  according to the
Frank-Condon principle).

In this paper we propose {\it the direct method} of
calculation of CF excitation energy in the framework of the LDA+U 
approximation which allows for such relaxation of 
electron system and enables to avoid ambiguity in the choice of basis set 
(as in projection or downfolding procedure) as well. CF excitations are 
calculated as difference between the ground state energy and the
total energy of an excited state in which the electron is
``artificially'' put to one of the higher-lying $d-$levels. 
These states - the ground state and the 
excited state with the electron constrained in the higher level - are 
both treated in the self-consisted LDA+U scheme. This allows us to obtain the values of 
CF excitation energy which, in particular, take into account fast relaxation of the electronic system.

Below we develop this general method and test it on the example of
Y and La titanates. 
We compare our results with the previously obtained ones.\cite{Pavarini-04,Solovyev-03} 
The physics of these compounds is discussed on the basis
of calculated values of CF excitation energy, as well as the other results of
{\it ab initio} LDA and LDA+U calculations.

\section{PHYSICS OF TITANATES AND CRYSTAL-FIELD SPLITTING}

Unusual and very rich physics of titanates attracts  much
attention since the possibility of existence of an orbital liquid
has been proposed for LaTiO$_3$.\cite{Keimer-00,Khaliullin-00} In
the ground state both LaTiO$_3$ and YTiO$_3$ crystallize in
perovskite structure.  The lattice distortions due to different
ionic radii of Y and La ions seems to lead to the cardinal changes 
in electronic and magnetic properties of these compounds.

At low temperatures LaTiO$_3$ was found to be G-type
antiferromagnet with N$\acute e$el temperature for stoichiometric
samples T$_N$=146~K~\cite{Cwik-03} while YTiO$_3$ is an isotropic
ferromagnet with relatively low T$_C\sim$30~K.\cite{Hester-97}
Local magnetic moment on Ti$^{3+}$ ion in YTiO$_3$ was found to be
0.84~$\mu_B$.\cite{Hester-97} The ordered moment in LaTiO$_3$
amounts to 0.46 - 0.58~$\mu_B$~\cite{Cwik-03,Keimer-00,Meijer-99}
and strongly differs from 1~$\mu_B$ expected for $d^1$
configuration with quenched orbital moment. Another interesting
feature is nearly isotropic magnon spectra with small spin gap
observed both in LaTiO$_3$ and YTiO$_3$ despite of the intrinsic
orthorhombic distortion.\cite{Keimer-00,Ulrich-02}

In order to describe these puzzling properties the exciting idea of the
orbital liquid has been proposed.\cite{Khaliullin-00} According to this
theory the large degeneracy of $t_{2g}$ shell can lead to the
orbitally-disordered ground state. The crucial point in this case is
the value of energy required for electron excitation from one $t_{2g}$
orbital to another: orbital liquid can exist only if this energy is zero or 
very small.

However, as the real symmetry of LaTiO$_3$ and YTiO$_3$ is not  cubic
but orthorhombic, one should expect certain splitting  of
$t_{2g}$-levels. First calculations of the CFS 
in LaTiO$_3$ was performed by R.
Schmitz and E. M\"uller-Hartmann.\cite{Cwik-03,Schmitz-04} 
Using realistic crystal structure\cite{Cwik-03} they carried out
model calculations taking into account point-charge and covalency contributions
and obtained that the lowest singlet
is separated from two higher-lying almost degenerate levels by
about 0.24 eV.  Similar calculations were also performed by
Mochizuki and Imada.\cite{Imada-03} They stressed the importance of
the GdFeO$_3$-type distortion and found the value of CFS to be
$0.77/\epsilon_{TiLa}$~eV, where $\epsilon_{TiLa}$ is an effective
dielectric constant. This constant is hard to evaluate in model
calculations in a reliable way because of local screening effects
in a solid, which could be explicitly taken into account only in
{\it ab initio} band structure calculations.

The {\it ab initio} calculations were done by Pavarini
{\it et al.}\cite{Pavarini-04} and Solovyev.\cite{Solovyev-03} 
They used the diagonalization of few orbital on-site effective LDA 
Hamiltonian in real space to obtain CFS. Solovyev has performed 
an exact procedure of full-orbital Hamiltonian transformation 
to the small energy-dependent Hamiltonian. After that the energy\vspace{-0.3cm}
\begin{center}
\begin{figure*}
 \centering
 \includegraphics[clip=false,width=0.42\textwidth]{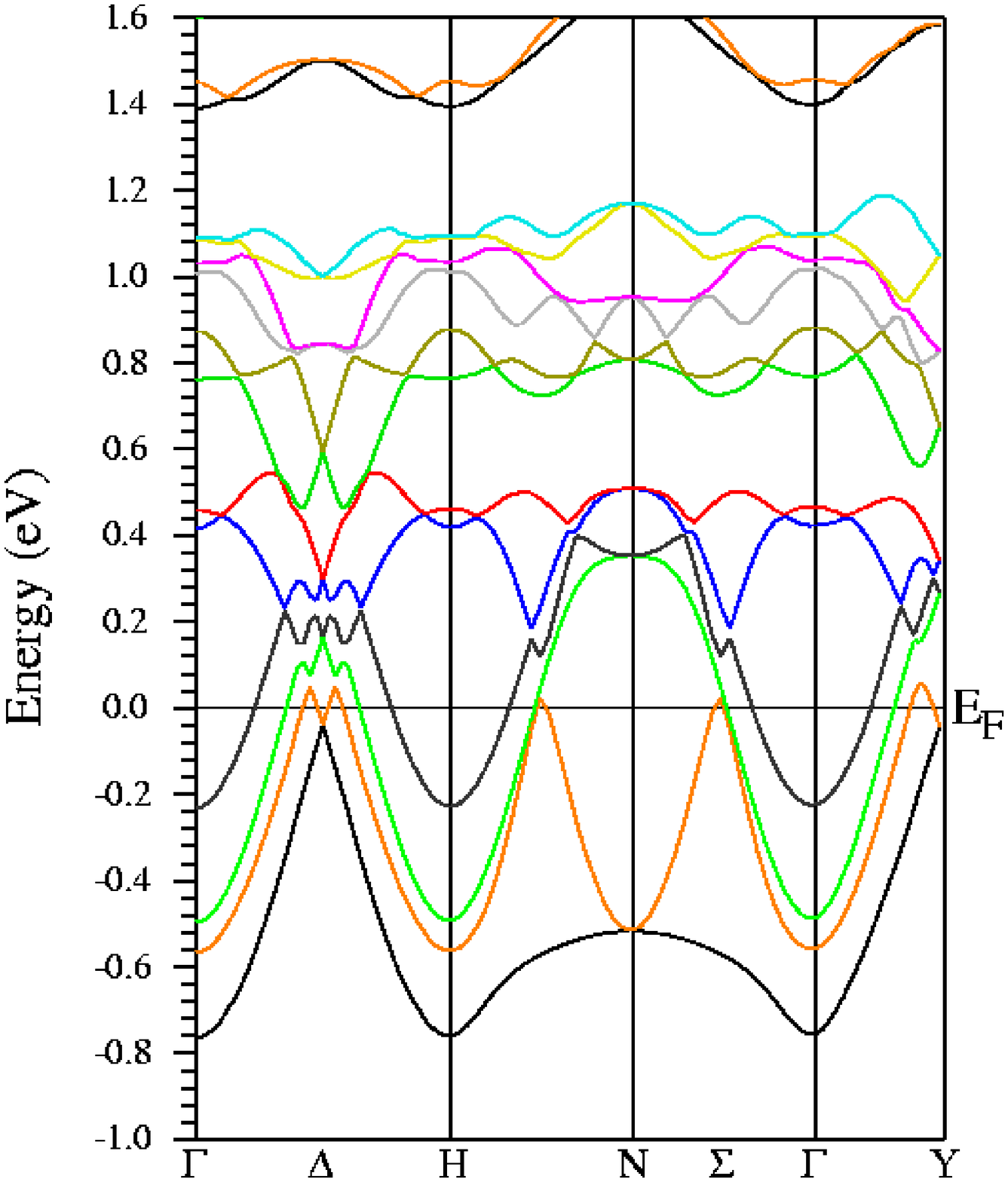}
 \includegraphics[clip=false,width=0.35\textwidth]{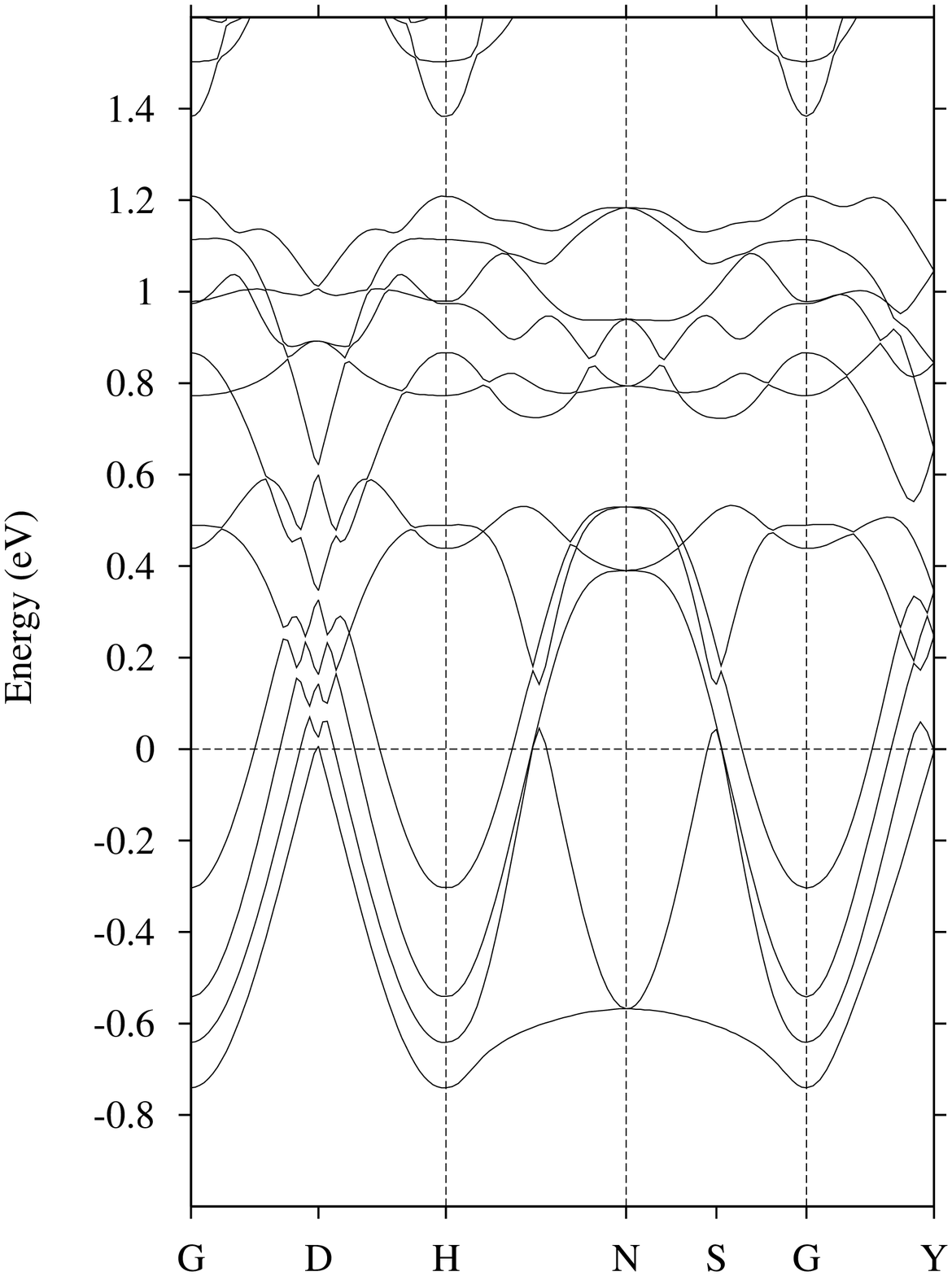}
\caption{\label{bands} (color online) Band structure for LaTiO$_3$ obtained within the LDA
approximation in the framework of two methods: the full potential
linearized augmented plane waves -- FP-LAPW (left) and linear
muffin-tin orbitals -- LMTO (right). The energy region is chosen
to illustrate the behavior of $t_{2g}$ bands. The Fermi level
corresponds to zero energy.} 
\end{figure*}
\end{center}\noindent was fixed in the center of the $t_{2g}$
band.\cite{Solovyev-03} Pavarini {\it et al.}\cite{Pavarini-04}
have used formalism of Nth-order muffin-tin
orbitals (NMTOs) to define small Hamiltonian.\cite{Andersen-00}
Despite the similarity of the methods,  the
results are qualitatively different. The CFS between lowest and
next $t_{2g}$ level obtained in Ref.~\onlinecite{Solovyev-03} is rather small, 
$\sim$30-50~meV for both compounds, which would not prevent
an orbital liquid formation. 
However, the results of calculations presented in Ref.~\onlinecite{Pavarini-04}
are incompatible with this scenario: obtained values of CFS
are $\sim$ 150-200~meV for both compounds, similar to the results of 
model calculations.\cite{Cwik-03,Imada-03}

\section{LDA: BAND STRUCTURE}
Discrepancy between the results of previous {\it ab initio} CFS
calculations~\cite{Pavarini-04,Solovyev-03} is probably caused by
(i) choice of MT spheres radii in LMTO method~\cite{Andersen-75} or
(ii) the details of different projection procedures. To
resolve first problem we have carried out the conventional LDA
calculations using LMTO method and verified its band structure by
performing the full-potential calculations in the framework of
linearized augmented plane waves (FP-LAPW) method realized on
Wien2k program code.\cite{Wien2k} We chose FP-LAPW  because it is
known to be the most accurate method for band structure
calculations. 

Crystallographic data for LaTiO$_3$ (T=8~K) and YTiO$_3$ (T=293~K) 
were taken from Ref.~\onlinecite{Cwik-03}.
The radii of the muffin-tin (MT) spheres for LMTO calculations 
were chosen to be R$_{\rm La }$=3.28~a.u., R$_{\rm Ti}$=2.52~a.u., and for
both oxygens R$_{\rm O}$=1.92~a.u. Remaining unit cell volume was
filled by empty spheres (atomic spheres with zero nuclear charge)
with different MT radii. Ionic radius of Y is known to be smaller than 
La one.\cite{Shannon} According to this fact
the MT radius of Y was also taken to be smaller: R$_{\rm Y }$=3.01~a.u.

The Ti($4s$,$4p$,$3d$), O($3s$,$2p$,$3d$) and
Y($5s$,$5p$,$4d$,4$f$), La($6s$,$6p$,$5d$) states were included
in the basis set in our calculations. Almost empty La-$4f$
states were treated as pseudo-core, since
La-$4f$ states are localized and do not strongly contribute to the
relevant electronic states. Due to rather
large values of the on-site Coulomb interaction $U$ (of the order
of 10~eV)  La-4f states will be located
approximately at 5-10~eV above the Fermi level.\cite{Chainani-92}

The Brillouin-zone (BZ) integration in the course of the
self-consistent iterations was performed over a mesh of 27 {\bf
k}-points in the irreducible part of the BZ.  Density of states (DOS)
were calculated by the tetrahedron method with 512 {\bf k}-points in
the whole BZ.

 The band structure obtained within the LDA approximation in the
framework of LMTO (with the MT radii chosen above) and FP-LAPW
methods for LaTiO$_3$ is presented in Fig.~\ref{bands}.  Twelve
Ti$-t_{2g}$ (four Ti per unit cell) bands are placed in the
vicinity of the Fermi level and have band width  
$\sim 1.95$ eV both in FP-LAPW and LMTO methods. The energy gap
between the top of the  $t_{2g}$ and the bottom of $e_g$ bands is
estimated to be 0.25~eV in FP-LAPW and  0.18~eV in LMTO. Comparing this figure 
and Fig.~2 of Ref.~\onlinecite{Solovyev-03} one can see
that the present LMTO bands agree better in all
high-symmetry points of BZ with FP-LAPW ones than those presented
in Ref.~\onlinecite{Solovyev-03}. So being firmly convinced of the
correctness of LMTO band structure one could carry out the CFS
calculations.

\section{LDA: WANNIER FUNCTION PROJECTION FOR CRYSTAL-FIELD SPLITTING CALCULATION}

One of the ways to calculate CFS is to use the minimal basis set Hamiltonian for
the orbitals of interest in real space. To construct such Hamiltonian we used
Wannier functions projection procedure. In this section we present 
brief description of the method. For more details, see 
Ref.~\onlinecite{Anisimov-04}. 
 
 For an LDA Hamiltonian
$\hat H$ one has a Hilbert space of eigenfunctions (Bloch states
$|\psi_{i\bf k}\rangle$) with the basis set $|\phi_\mu\rangle$
defined by the particular method (for LMTO these are
LMT-orbitals,\cite{Andersen-75} for LAPW - Augmented Plane
Waves,\cite{Mattheiss86} etc.). In this basis set the Hamiltonian
operator is defined as:
\begin{eqnarray}
\label{Ham_gen}
 \widehat H & = & \sum_{\mu\nu} |\phi_\mu\rangle
H_{\mu\nu} \langle\phi_\nu|,
\end{eqnarray}
where greek indices are used for full-orbital matrices.

If we consider a certain subset of the Hamiltonian eigenfunctions,
for example Bloch states of partially filled bands $|\psi_{n\bf
k}\rangle$, then we can define a corresponding subspace in the
total Hilbert space. The Hamiltonian matrix is diagonal in the
Bloch states basis, however, physically more appealing is a basis which
would have a form of  site-centered atomic orbitals. That is a set
of Wannier functions $|W_{n}^{\bf T}\rangle$ defined as a Fourier
transformation of certain linear combination of Bloch functions
belonging to this subspace (see Eq. (\ref{WF_psi_def}) bellow). They
are labeled in real space according to the band $n$ and the
lattice vector of the unit cell {\bf T} which they belong to. The
Hamiltonian operator $\hat H^{WF}$ defined in this basis set is
\begin{eqnarray}
\label{Ham_genWF} \widehat H^{WF} & = & \sum_{nn'{\bf
T}}|W_{n}^{\bf 0}\rangle H_{nn'}({\bf T}) \langle W_{n'}^{\bf T}|.
\end{eqnarray}

The total Hilbert space can be divided into a direct sum of above
introduced subspace (of partially filled Bloch states)
and the subspace formed by all other states orthogonal to it.
Those two subspaces are decoupled since they are the eigenfunctions
corresponding to different eigenvalues. 
The Hamiltonian matrix in Wannier function basis is block-diagonal, so
that the matrix elements between different subspaces in Hilbert
space are zero. The block $H_{nn'}$ in (\ref{Ham_genWF})
corresponding to the partially filled bands can be considered as a
projection of the full Hamiltonian operator (\ref{Ham_gen}) to the
subspace defined by its Wannier functions.

Localized Wannier functions $|W_{i}^{\bf T}\rangle$ were defined
in Ref.~\onlinecite{wannier} as Fourier transforms of the Bloch functions
$|\psi_{i\bf k}\rangle$
\begin{eqnarray}
\label{WF_psi_def}
|W_i^{\bf T}\rangle & = & \frac{1}{\sqrt{N}} \sum_{\bf k}
e^{-i{\bf kT}}|\psi_{i{\bf k}}\rangle,
\end{eqnarray}
where $N$ is the number of discrete $\bf k$ points in the
first BZ.

Wannier functions are not uniquely defined because for a certain
set of bands any orthogonal linear combination of Bloch functions
$|\psi_{i\bf k}\rangle$ can be used in (\ref{WF_psi_def}). In
general it means that the freedom of choice of Wannier functions
corresponds to  freedom of choice of a {\bf k}-depended unitary transformation
matrix $U^{({\bf k})}_{ji}$:\cite{vanderbildt}
\begin{eqnarray}
\label{psi_def}
|\widetilde\psi_{i\bf k}\rangle & = & \sum_j
U^{({\bf k})}_{ji} |\psi_{j\bf k}\rangle.
\end{eqnarray}

The most serious drawback of Wannier representation is that there is no rigorous
way to define $U^{({\bf k})}_{ji}$. As one can see from
(\ref{WF_psi_def}) and (\ref{psi_def}), Wannier functions can vary
significantly in shape and range because variations in 
$|\psi_{i\bf k}\rangle$ or $U^{({\bf k})}_{ji}$ change relative phases and 
amplitudes of Bloch functions at different $\bf k$ and bands $i$. 

In order to avoid this disadvantage some additional restrictions on the 
properties of Wannier functions have been proposed.\cite{Koster-53,Kohn-73,vanderbildt}
Among others Marzari and
Vanderbildt~\cite{vanderbildt} suggested the condition of maximum
localization for Wannier functions. That gave the variational
procedure to calculate $U^{({\bf k})}_{ji}$. To get a good initial
point the authors suggested to choose a set of trial localized
orbitals $|\phi_n\rangle$ and projecting them onto the Bloch
functions $|\psi_{i\bf k}\rangle$. It was found that this starting
guess is usually quite good.\cite{vanderbildt} This fact led
later to the simplified calculating scheme proposed in
\cite{pickett} where variational procedure was abandoned and the
result of aforementioned projection was considered as a final
step.

In order to start projection procedure one needs to determine
the set of trial orbitals $|\phi_n\rangle$ and the bands which will be
used for the Wannier functions construction. The latter can be
defined either by the bands numbers (from $N_1$ to $N_2$) or by the
energy interval ($E_1, E_2$).

Non-orthogonalized Wannier functions in real
$|\widetilde{W}_{n}^{\bf T}\rangle$ and reciprocal space
$|\widetilde{W}_{n\bf k}\rangle$ are then the projection of the
set of trial site-centered atomic-like orbitals $|\phi_n\rangle$
on the Bloch functions $|\psi_{i\bf k}\rangle$
of the chosen bands
\begin{eqnarray}
\label{WF_psi} |\widetilde{W}_{n}^{\bf T}\rangle & = &
\frac{1}{\sqrt{N}} \sum_{\bf
k} e^{-i{\bf kT}} |\widetilde{W}_{n\bf k}\rangle, \\ \nonumber
|\widetilde{W}_{n\bf k}\rangle & \equiv & \sum_{i=N_1}^{N_2}
|\psi_{i\bf k}\rangle\langle\psi_{i\bf k}|\phi_n\rangle =
\sum_{i(E_1\le \varepsilon_{i}({\bf k})\le E_2)} |\psi_{i\bf
k}\rangle\langle\psi_{i\bf k}|\phi_n\rangle.
\end{eqnarray}

Note that the Wannier functions in reciprocal space
$|\widetilde{W}_{n\bf k}\rangle$ do not coincide with the Bloch
functions $|\psi_{n\bf k}\rangle$ for multi-band case due to the
summation over band index $i$ in (\ref{WF_psi}). One can consider
them as Bloch sums of Wannier functions analogous to the basis
functions Bloch sums $\phi_j^{\bf k}({\bf r})$, see (\ref{psik})
below.

\begin{table*}
\centering \caption{The magnitudes of CFS in $t_{2g}$ shell (in
meV) obtained in the LDA approach by different authors are
presented in first 3 columns. First value is an energy difference
between the lowest energy level and the middle one; the second is
the difference between the middle and the highest energy
levels in $t_{2g}$ shell. The results of the constrained LDA+U
calculations, which take into account fast electron relaxations 
are shown in the last column. } \vspace{0.2cm} \label{SplitTable}
\begin{tabular}{ccccc}
\hline
\hline
 & Present results & E. Pavarini {\it et al.}~\footnote{Reference \cite{Pavarini-04}} & I.V.
 Solovyev~\footnote{Reference \cite{Solovyev-03}} & Present results\\
 & (WF projection) &  &  & (LDA+U)\\
\hline
$LaTiO_3$ & 230; 40 & 140; 60  & 54; 39  & 160; -\\
$YTiO_3$  & 180; 80 & 200; 130 & 27; 154 & 150; -\\
\hline
\end{tabular}
\end{table*}

The coefficients $\langle\psi_{i\bf k}|\phi_n\rangle$ in
(\ref{WF_psi}) define (after orthonormalization) the unitary
transformation matrix $U^{({\bf k})}_{ji}$ in (\ref{psi_def}).
However, projection procedure defined in (\ref{WF_psi}) can be
considered as a more general formalism than unitary transformation
(\ref{psi_def}).

The Bloch functions in LMTO basis take the form
\begin{eqnarray}
\label{psi} |\psi_{i\bf k}\rangle=
\sum_{\mu} c^{{\bf k}}_{\mu i}|\phi_{\mu}^{\bf k}\rangle,
\end{eqnarray}
where $\mu$  is the combined index of the $qlm$ ($q$ - atomic number
in the unit cell, $lm$ are orbital and magnetic quantum numbers),
and $\phi_{\mu}^{\bf k}({\bf r})$ are Bloch sums of the basis orbitals
$\phi_{\mu}({\bf r- T})$
\begin{eqnarray}
\label{psik} \phi_{\mu}^{\bf
k}({\bf r}) & = & \frac{1}{\sqrt{N}} \sum_T e^{ikT} \phi_{\mu}({\bf r}-{\bf T}),
\end{eqnarray}

\noindent and the coefficients have the property
$c^{\bf k}_{\mu i} = \langle\phi_{\mu}|\psi_{i\bf k}\rangle.$

If $n$ in $|\phi_n\rangle$ corresponds to the particular $qlm$
combination (in other words $|\phi_n\rangle$ is an {\it orthogonal} LMTO
basis set orbital), then $\langle\psi_{i\bf k}|\phi_n\rangle =
c_{ni}^{{\bf k}*}$ and hence 
\begin{eqnarray}
\label{WF} 
|\widetilde{W}_{n\bf
k}\rangle & = &  \sum_{i=N_1}^{N_2} |\psi_{i\bf k}\rangle
c_{ni}^{{\bf k}*} \\ \nonumber & = &
 \sum_{i=N_1}^{N_2} \sum_{\mu} c_{\mu i}^{\bf k} c_{ni}^{{\bf k}*}
|\phi_{\mu}^{\bf k}\rangle = 
\sum_{\mu} \tilde{b}^{\bf k}_{\mu n}
|\phi_{\mu}^{\bf k}\rangle,
\end{eqnarray}
\begin{eqnarray}
\nonumber
\tilde{b}^{\bf k}_{\mu n} =\sum_{i=N_1}^{N_2} c_{\mu i}^{\bf k} c_{ni}^{{\bf k}*}.
\end{eqnarray}

In order to orthonormalize the WF (\ref{WF}) an overlap matrix
$O_{nn'}({\bf k})$ and its inverse square root $S_{nn'}({\bf k})$
can be defined as
\begin{eqnarray}
\label{O-S} O_{nn'}({\bf k})&\equiv& \langle\widetilde{W}_{n\bf
k}|\widetilde{W}_{n'\bf k}\rangle = \sum_{i=N_1}^{N_2} c_{ni}^{\bf
k} c_{n'i}^{{\bf k}*}, \\ \nonumber S_{nn'}({\bf k})
&\equiv& O^{-1/2}_{nn'}({\bf k}),
\end{eqnarray}
The orthogonality of Bloch states $\langle\psi_{n\bf
k}|\psi_{n'\bf k}\rangle=\delta_{nn'}$ was used.

Orthonormalized Wannier functions in $k$-space $|W_{n\bf
k}\rangle$ can be obtained as
\begin{eqnarray}
\label{WF_orth} |W_{n\bf k}\rangle & = &\sum_{n'} S_{nn'}({\bf k})
|\widetilde{W}_{n'\bf k}\rangle =
\sum_{i=N_1}^{N_2} |\psi_{i\bf k}\rangle \bar{c}_{ni}^{{\bf
k}*}
\\ \nonumber &=&\sum_{\mu} b^{\bf k}_{\mu n} |\phi_{\mu}^{\bf k}\rangle,
\end{eqnarray}

where
\begin{eqnarray}
\label{other} 
\bar{c}_{ni}^{{\bf k}*}&\equiv& \langle\psi_{i {\bf k}}|W_{n {\bf k}}\rangle=
\sum_{n'} S_{nn'}({\bf k})
c_{n'i}^{{\bf k}*},
\\ b^{\bf k}_{\mu n} &\equiv& \langle\phi_{\mu}^{\bf k}|W_{n {\bf k}}\rangle=
\sum_{i=N_1}^{N_2} \bar{c}_{\mu i}^{\bf k} \bar{c}_{ni}^{{\bf k}*}.
\end{eqnarray}

Thus, matrix elements of the few orbital Hamiltonian $\widehat H^{WF}$  in the basis
of WF in real space where both orbitals are in the same unit cell
are given by the following expression
\begin{eqnarray}
\label{E_WF} H^{WF}_{nm}(0) & = & \langle W_{n}^{\bf
0}|\frac{1}{N}\biggl(\sum_{\bf k}\sum_{i=N_1}^{N_2}|\psi_{i\bf k}\rangle
\epsilon_{i}({\bf k})\langle\psi_{i\bf k}|\biggr)|W_{m}^{\bf
0}\rangle = \nonumber \\
       & = & \frac{1}{N} \sum_{\bf k}\sum_{i=N_1}^{N_2} \bar{c}_{ni}({\bf k})
      \bar{c}_{mi}^{*}({\bf k})\epsilon_{i}({\bf k}).
\end{eqnarray}
Here $\epsilon_{i}({\bf k})$ is an eigenvalue for a particular band.

For both LaTiO$_3$ and YTiO$_3$ we are interested in CFS in the
$t_{2g}$ shell. For that 3$\times$3 Hamiltonians were
derived for $t_{2g}$ bands placed in the vicinity of the Fermi
level in the energy range (-0.8; 1.25)~eV  for LaTiO$_3$ and 
(-0.65; 1.55)~eV for YTiO$_3$. Then we diagonalize them and
calculate the difference between corresponding eigenvalues in
order to determine the CFS.

The obtained values of CFS together with results of 
Ref.~\onlinecite{Pavarini-04,Solovyev-03} are presented in Tab.~\ref{SplitTable}. 
The CFS calculated by Pavarini {\it et al.}\cite{Pavarini-04} has the same 
character and similar values as ours. 
According to these results, in LaTiO$_3$ the lowest energy level is
widely separated from two other, which are almost degenerate. 
Similar result was obtained by Cwik {\it et al.} using  a full 
Madelung-sum point charge model, where the first splitting was found 
to be 240~meV.\cite{Cwik-03}

The situation is a little bit different in YTiO$_3$.  The
magnitude of the first splitting  (between lowest and middle
energy levels) has the same order as in La titanate, 
while the second one (between middle and highest
energy levels) increases. It is interesting to note that in
calculations of Solovyev the same effect of the second
splitting enhancement going from LaTiO$_3$ to YTiO$_3$ is
observed.\cite{Solovyev-03} Nevertheless, the character of CFS
obtained by Solovyev~\cite{Solovyev-03} is quite
different from the present one and from that calculated in
Ref.~\onlinecite{Pavarini-04}.

\section{LDA+U: BAND STRUCTURE}

The application of LDA to solids provides important information about details
of band structure. However, for transition metal compounds it
usually leads to metallic type of the electronic structure due to
the presence of the partially filled $d-$bands at the Fermi level,
often in contrast with an experiment. This is also the
case in the LDA approach presented above (Sec. IV). The LDA+U
approximation has been developed to include in the calculation scheme  
orbital-dependent Hubbard-like correction $U$ which acts differently 
on the occupied and unoccupied $d-$orbitals giving as a result 
correct description for localized states.\cite{Anisimov-91}

In order to analyze of the results of the LDA+U
calculations in terms of $t_{2g}$ and $e_{g}$ orbitals
one has to chose in the local coordinate system (LCS). Usually
it is possible to define the LCS where axes are pointed directly
along $Ti-O$ bonds, but for strongly distorted compounds
(like LaTiO$_3$ and YTiO$_3$) such coordinate system in general might be
not rectangular. We use the rectangular local coordinate system where the
axes are directed as much as possible to the nearest oxygens.

The interorbital on-site Coulomb interaction parameter $U$ and 
intra-atomic exchange coupling $J$ for $t_{2g}$ shell were estimated 
to be 3.3~eV and 0.8~eV, respectively, using constrained superscell
calculations~\cite{Pickett-98} in the framework of LMTO method and
taking into account screening by $e_g$ electrons.\cite{Solovyev-96} 
These values are in good agreement with previous
findings.\cite{Solovyev-96,Mizokawa-96} 
The experimentally observed magnetic structures: YTiO$_3$ -
ferromagnet, LaTiO$_3$ - G-type antiferromagnet are used 
for the conventional and constrained LDA+U calculations.

\begin{center}
\begin{figure}
 \centering
 \includegraphics[clip=false,width=0.42\textwidth]{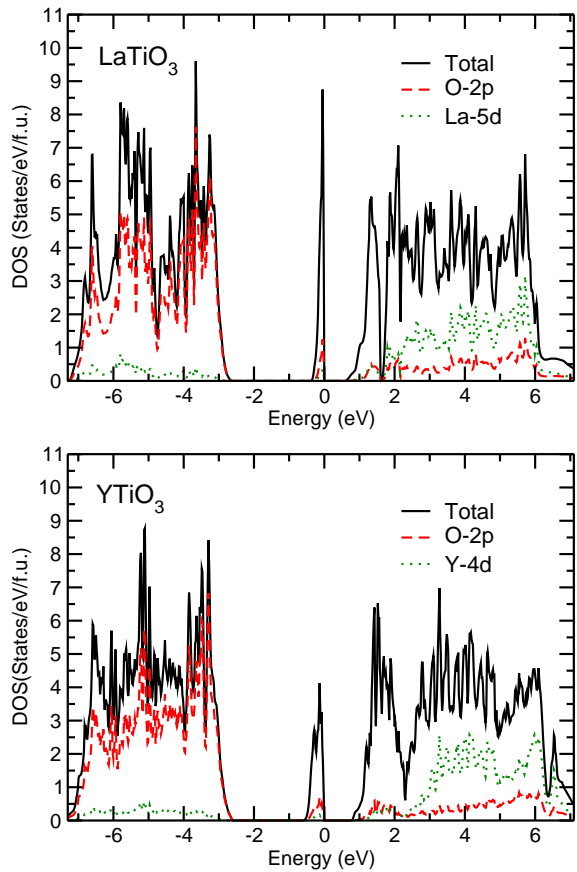}
\caption{\label{LDAU-DOS}(color online). Total and partial DOS of LaTiO$_3$ (top panel) and
YTiO$_3$ (bottom panel) calculated within the LDA+U approximation for
AFM-G and FM magnetic structures, respectively. The Fermi level corresponds to zero energy.}
\end{figure}
\end{center}

The top panel of Fig.~\ref{LDAU-DOS} shows LDA+U 
electronic structure of LaTiO$_3$ in details.  There are three
distinguishable sets of bands: completely filled O-$2p$ bands,
partially filled Ti-$3d$ bands and empty La-$5d$ bands.   The bands in
the energy range from -7.2~eV to -2.7~eV originate mainly  from O-2$p$
states. The gap $\sim$2.3~eV appears between O-$2p$ and the narrow
peak of Ti-$3d$($t_{2g}$) states. The band gap in LaTiO$_3$ is found to
be 0.57~eV. It divides Ti-$3d(t_{2g})$ band into two parts in such a way
that both the top of the valence and the bottom of the conduction band are
predominantly formed by Ti-$3d(t_{2g})$ states (see Fig.~\ref{LaTiO3-d-DOS}). 
The bands lying higher than $\sim$2~eV have
basically La-$3d$, O-$2p$ and Ti-$3d(e_g)$ contributions.
The calculated local magnetic moment on Ti$^{3+}$ ion amounts to 0.78$\mu_B$
being by $\sim$0.2$\mu_B$ larger than the experimentally observed value.
\cite{Cwik-03}
\vspace{-0.2cm}

The electronic structure obtained in the LDA+U approximation 
for YTiO$_3$ is shown in the bottom panel of Fig.~\ref{LDAU-DOS}. 
Generally it has similar character.
However, the value of the band gap is larger than in LaTiO$_3$ 
($\sim$0.78~eV). The value of the local magnetic moment is found 
to be 0.89$\mu_B$ in good agreement with experiment.\cite{Hester-97}

There are two opposite factors influencing the magnitude of the band gap.
On one hand the larger lattice distortions in YTiO$_3$ due to smaller
ionic radii of Y in comparison with La make Ti($3d$)-O($2p$)
hybridization weaker and as a result the band gap larger. On the contrary,
ferromagnetism observed in  YTiO$_3$ leads to an increase of the band
width resulting in a band gap reduction.

\begin{table*}
\centering
\caption{Total energy difference between three magnetic
solutions per Ti ion and magnitudes of the band gaps and local magnetic
moments on Ti ions obtained in LDA+U calculation. Zero energy is the
lowest total energy for compound under consideration.}
\vspace{0.2cm}
\label{EnergyTable}
\begin{tabular}{l|ccc|ccc}
\hline\hline
\multicolumn {1}{c}{}& \multicolumn {3}{c}{$LaTiO_3$} & \multicolumn
{3}{c}
{$YTiO_3$} \\
\multicolumn {1}{c}{}&
\multicolumn {1}{c}{Band gap}&
\multicolumn {1}{c}{Mag. moment}&
\multicolumn {1}{c}{Total energy}
&
\multicolumn {1}{c}{Band gap}&
\multicolumn {1}{c}{Mag. moment}&
\multicolumn {1}{c}{Total energy}\\
\hline
\multicolumn {1}{c}{FM}&
\multicolumn {1}{c}{0.45~eV}&
\multicolumn {1}{c}{0.88~$\mu_B$}&
\multicolumn {1}{c}{125~K}
&
\multicolumn {1}{c}{0.78~eV}&
\multicolumn {1}{c}{0.89~$\mu_B$}&
\multicolumn {1}{c}{0}

\\
\multicolumn {1}{c}{AFM-A}&
\multicolumn {1}{c}{0.54~eV}&
\multicolumn {1}{c}{0.83~$\mu_B$}&
\multicolumn {1}{c}{0}
&
\multicolumn {1}{c}{0.89~eV}&
\multicolumn {1}{c}{0.87~$\mu_B$}&
\multicolumn {1}{c}{60~K} \\
\multicolumn {1}{c}{AFM-G}&
\multicolumn {1}{c}{0.57~eV}&
\multicolumn {1}{c}{0.78~$\mu_B$}&
\multicolumn {1}{c}{40~K}
&
\multicolumn {1}{c}{1.04~eV}&
\multicolumn {1}{c}{0.81~$\mu_B$}&
\multicolumn {1}{c}{230~K} \\
\hline
\end{tabular}
\end{table*}

In order to clarify the role of magnetism we performed additional 
band structure calculations for magnetic structures different 
from experimental one. 

The FM, AFM-G and AFM-A structures 
were considered. The values of the band gap and local magnetic moments on
Ti$^{3+}$ ions for all calculated configurations are presented in Tab.
\ref{EnergyTable}.  One can see that the band gap
reduction due to ferromagnetism amounts to 21$\%
$ (0.12~eV) in the case of LaTiO$_3$ and 25$\%
$ (0.26~eV) in YTiO$_3$. The modifications of an
electronic structure with the change of magnetic structure
from AFM-G to FM in the vicinity of Fermi level for LaTiO$_3$ are presented
in the inset of Fig.\ref{LaTiO3-d-DOS}.

Continuing investigation of the interplay between lattice, electronic and
magnetic degrees of freedom one can isolate the influence of ``pure lattice
distortions'' on the electronic structure of La and Y titanates comparing
the results of calculations performed in identical magnetic structures.
The analysis shows significant changes in the value of the band gap. Only
due to the lattice distortions it changes by 0.32-0.47~eV depending on the
magnetic structure under consideration.

 The main reason for such a
strong influence of local geometry on the electronic structure of these
compounds is probably connected with the stronger covalency of $d^1$
configuration and hence higher sensibility to the distortions  in
comparison with other configurations of $d-$shell (with the exclusion of $d^9$). 
In contrast to the band gap, magnetic
moment is only reduced by 10$\%$ with the change of magnetic configuration.

In addition, it should be mentioned that our calculations 
unlike the experiment give for LaTiO$_3$ the lowest total energy 
for A-type AFM, although experimentally observed G-type lies quite close. 
This result is supported by the calculation of the exchange interaction 
parameters~\cite{Solovyev-03} and connected with the orbital pattern of the compound 
discussed in details in Sec.~\ref{OS}.

\section{CALCULATION OF CRYSTAL-FIELD EXCITATION ENERGY IN 
 LDA+U}

The direct study of CFS, for instance by optical measurements, implies the
excitation of an electron from one energy level to another.
Simultaneously with this excitation the external ``bath'' formed by
the rest of the electrons can relax giving the system a chance
to lower the energy. The simple CFS computations using the centers of 
gravity of corresponding bands, projection or downfolding procedure 
do not take into account such processes and as a result overestimate the
value of CFS. In this section we propose 
the direct calculation of the total energy difference between ground
and first excited states ({\it CF excitation energy calculation}) 
in $t_{2g}$ shell using
the constrain procedure adopted for the LDA+U approximation.
\begin{center}
\begin{figure}[b]
 \centering
 \includegraphics[clip=false,angle=270,width=0.4\textwidth]{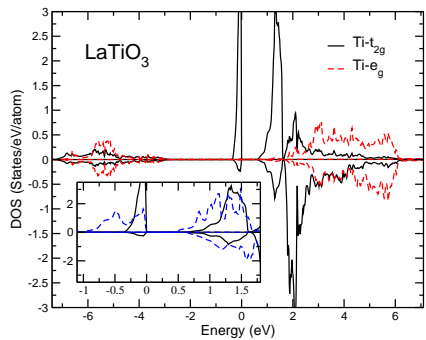}
\caption{\label{LaTiO3-d-DOS}(color online). LaTiO$_3$. Ti-$3d$ partial 
DOS calculated within LDA+U approach.
The inset shows Ti-$3d$ partial DOS in AFM-G (solid line) and FM
(dashed line) configurations. Parts of plots with positive (negative)
ordinates denote majority (minority) spin DOS.
The Fermi level corresponds to zero energy.}
\end{figure}
\end{center}

The occupation matrix in LDA+U can be defined in the usual way:
\begin{equation}
\label{eq:Occ}
  n_{mm'}^\sigma =-\frac 1\pi\int^{E_F}ImG_{inlm,inlm'}^\sigma(E)dE,
\end{equation}
where $\sigma$ is the spin, $i$ denotes the site, $n,l,m$ are principle,
orbital and magnetic
quantum numbers respectively, $G_{inlm,inlm'}^\sigma(E)=
\langle inlm\sigma|(E-\widehat H^{LDA+U})^{-1}| inlm'\sigma \rangle$ are
the elements of the Green function matrix and $\widehat H^{LDA+U}$ is
a single particle LDA+U Hamiltonian (for its definition see
Ref.~\onlinecite{Anisimov-91}). The occupation matrix for an isolated ion
is diagonal. However, for ions in solids it can have more complex 
structure in low symmetry systems and one needs to diagonalize it. 
The eigenvectors of the \eqref{eq:Occ} can be used for transformation 
of occupation matrix to diagonal form. Using this procedure one can 
obtain the information on the orbitals where the electrons are localized.
\begin{center}
\begin{figure*}
 \centering
 \includegraphics[clip=true,width=0.49\textwidth]{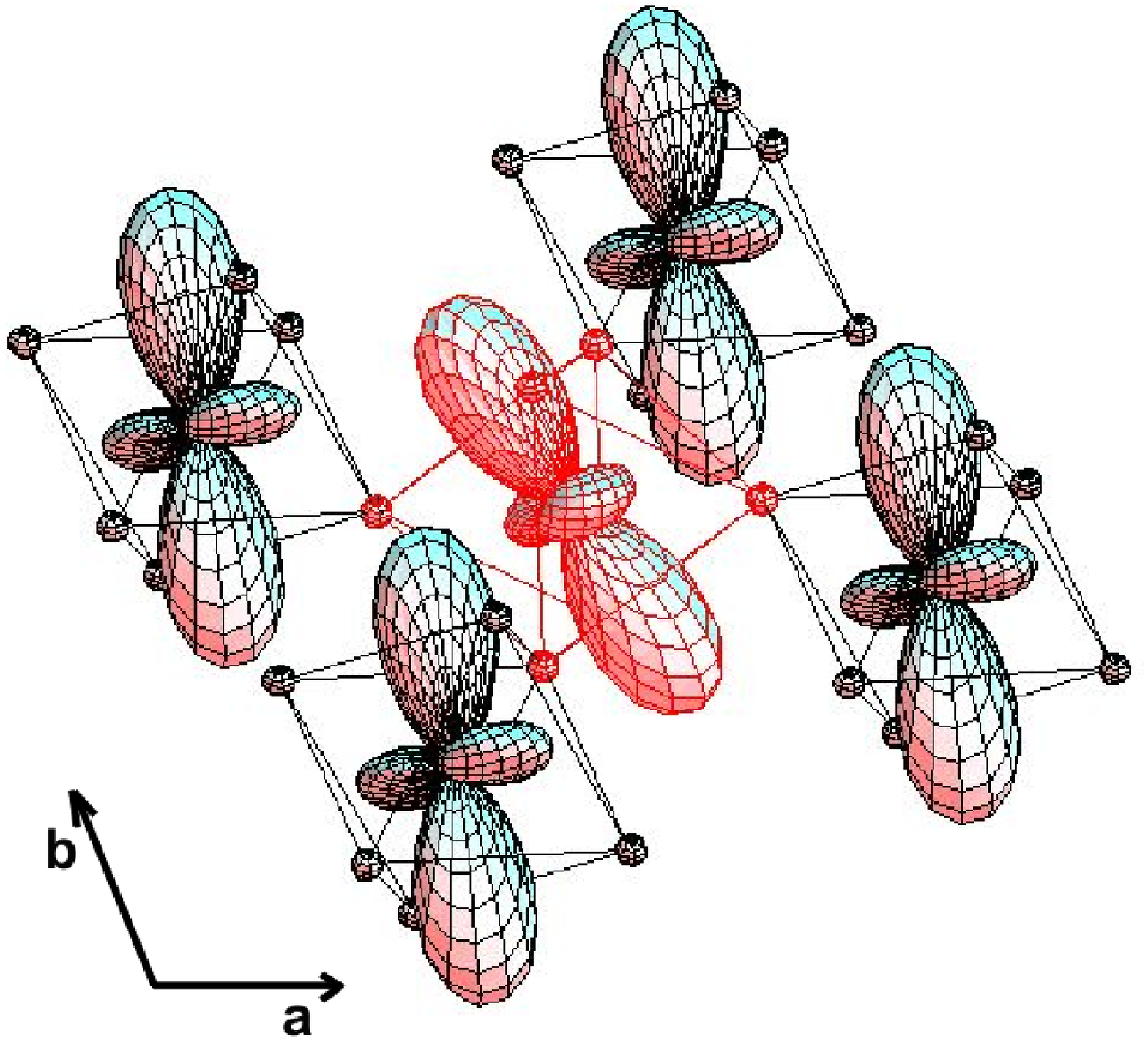}
 \includegraphics[clip=true,width=0.49\textwidth]{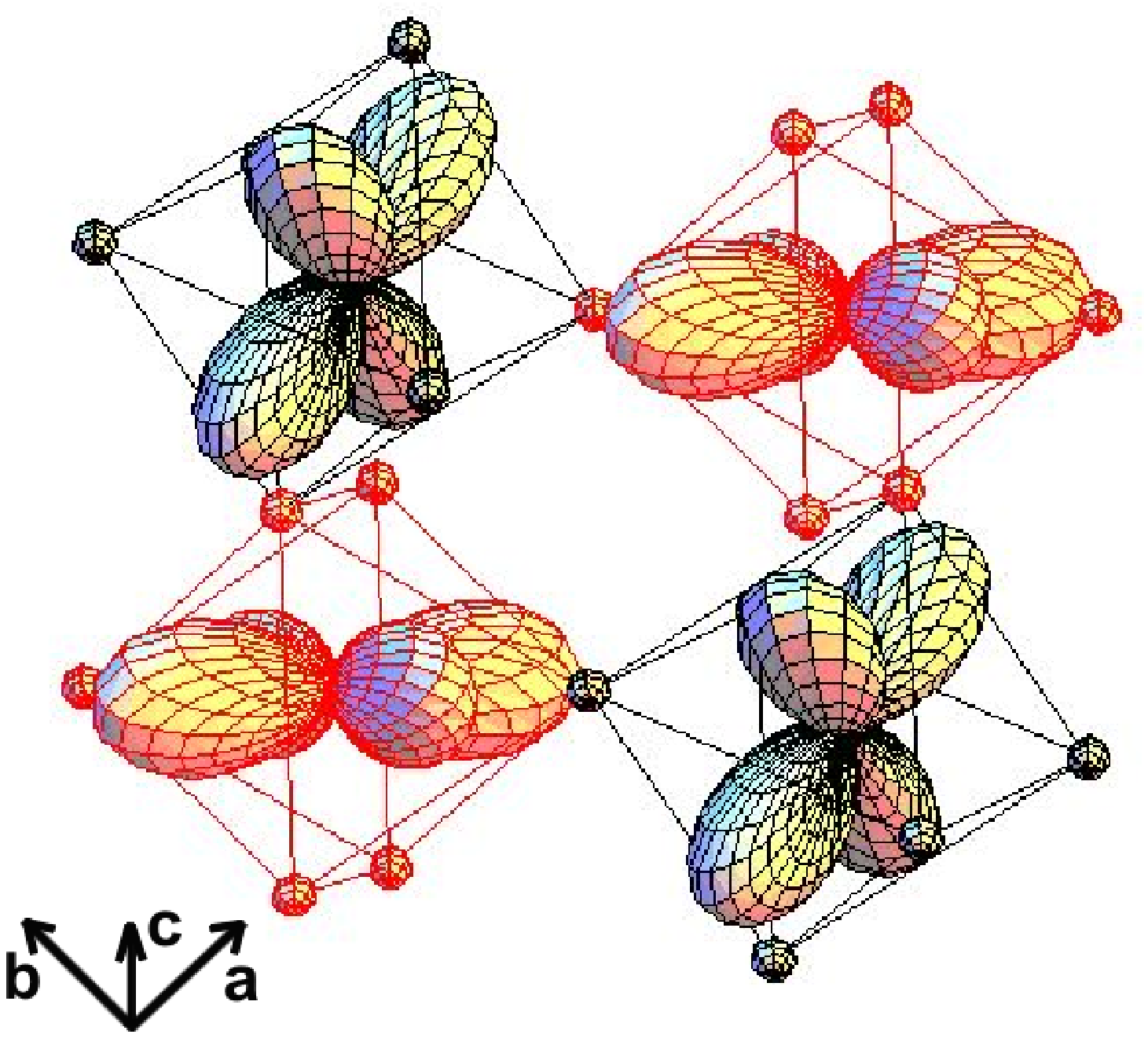}
\caption{\label{ab-plane}(color online) Orbital ordering in LaTiO$_3$ (left) and YTiO$_3$ (right) in
$ab$-plane obtained in the framework of the LDA+U calculations for AFM-G
and FM structures respectively.}
\end{figure*}
\end{center}

\begin{table}[b]
\centering
\caption{Decomposition of the lowest in energy orbital
into $xy$, $xz$ and $yz$ orbitals. The results are
presented in local coordinate system where the orthogonal 
axes point as much as possible to neighboring oxygens.}
\vspace{0.2cm}
\begin{tabular}{lccc}
\hline
\hline
& $LaTiO_3$ & $YTiO_3$ \\
\hline
LDA      & 0.63$xy$+0.66$yz$+0.40$xz$& 
0.76$xy$+0.64$yz$-0.06$xz$\\
LDA+U    & 0.62$xy$+0.72$yz$+0.32$xz$& 
0.75$xy$+0.66$yz$-0.04$xz$\\
\hline
\label{orb-table}
\end{tabular}
\end{table}

In fact, in the case of spin-polarized calculations (like LDA+U) there
are two occupation matrices for every transition metal ion on each
site: spin majority and spin minority occupation matrices. According to
the Hund's rule, at first the spin majority states start to be occupied.
Nevertheless, due to the large spatial extension and sizable overlap between
$e_g$ and oxygen $p$ orbitals (see Fig.~\ref{LaTiO3-d-DOS}) 
there are always some non-negligible occupation numbers for
$e_g$ states in a real band structure calculation for both
spins. In order to avoid the influence of these effects 
(they can lead to the ``symmetrization'' of $d$-orbitals) we use for the 
analysis not the occupation matrix for majority
spin (Ti is $d^1$ system in the compounds under consideration), but
the difference between matrices for two spins. 
Here we suppose the oxygen influence on $d-$states for both 
spins to be equal.

The diagonalization of the difference between occupation matrices for
majority and minority spins for LaTiO$_3$ gives following
eigenvalues:  0, 0, 0.01, 0.02, 0.76. The eigenvector corresponding to the
largest eigenvalue defines the occupied orbital as a linear combination of
all $d-$orbitals:
\begin{equation}
|\Psi_{GS}\rangle = \sum_{i} a_i |\psi_i \rangle.
\end{equation}
Thus, in case of LaTiO$_3$ in the LCS:
\begin{eqnarray}
\label{LaTiO3-occ-orb}
|\Psi^{LaTiO_3}_{GS}\rangle  = 0.62xy+0.72yz+0.32xz+\\ \nonumber
0.02(3y^2-r^2)+0.02(z^2-x^2).
\end{eqnarray}
Similar decomposition for YTiO$_3$ is presented in the Table III.

The excitation energy (from the ground to the first excited state) 
in this case is the total energy difference between states where 
this orbital is occupied and where  it
is empty and another one is occupied. According to this scheme we 
performed the calculation where the system is constrained by
external potential $\widehat V_{constr}$ to change the occupied orbital:
\begin{equation}
  \widehat V_{constr} = |\Phi_{GS} \rangle \delta V \langle \Phi_{GS}|.
\end{equation}
In other words $\widehat V_{constr}$ just pushes up the orbital 
where the electron has been localized in LDA+U. 
It is not important how big is the correction 
$\delta V$.  One needs just be sure that it is large enough 
to force the electron to hop to another orbital.
The total energy of excited state does not depend on
the value of $\delta V$, because the correction is applied 
to the orbital which has to be empty in the excited state.
The total energy difference between 
the ground and first excited states is an estimate of the CF excitation energy.

The results of calculations of CF excitation energies for LaTiO$_3$ and
YTiO$_3$ are presented in the fourth
column of Tab.~\ref{SplitTable}. 
They have the same order of magnitude
as those obtained using the WF projection in the present
work and downfolding performed by Pavarini {\it et al.}\cite{Pavarini-04}
As has been mentioned above, for proper estimation of CF excitation
energy the relaxation of electron system should be taken into
consideration. Such relaxation decreases the energy costs
of the electron excitation. According to the present
results the gain in energy can amount to $\sim$70~meV ($\sim$30\%).
However, it is obviously insufficient to reduce the value of CFS
significantly.

Thus, the results of the CFS calculations both using WF projection and
constrained LDA+U which takes into account the electron system relaxation
indicate that the splittings in LaTiO$_3$ and YTiO$_3$ are relatively large.
They are definitely bigger than the spin-orbital coupling in titanates
which is expected to be $\Lambda_{SO}\sim20$~meV~\cite{Abragam}.
The latter result is supported by the recent XAS measurements, where 
shown that the orbital momentum in LaTiO$_3$ is essentially 
quenched.\cite{Haverkort-04}

But the most important conclusion is that CF excitation energy 
obtained in the present work and value of CFS calculated
in Ref.~\onlinecite{Pavarini-04} as well as extracted from 
Ti L$_{2,3}$ XAS~\cite{Haverkort-04} and optical 
measurements~\cite{Gruninger-05} are all the order of 200 meV, 
that makes orbital liquid scenario for LaTiO$_3$ rather unlikely.

\section{ORBITAL STRUCTURE}
\label{OS}
Orbital degrees of freedom are known to play an important role
and should be correctly taken into account in theories describing
magnetic interactions in strongly correlated materials.\cite{Kugel-82}

As one can see from Tab. \ref{orb-table} the composition of the 
occupied orbitals in Y and La titanates is quite different. Contribution 
of $xy$ and $yz$ components to the resulting orbital in LaTiO$_3$ are 
nearly the same and by $\sim$~40--60 $\%
$ bigger than $xz$ one. There is quite different
situation in YTiO$_3$, where $xz$ contribution is almost zero.

This fact can be explained by means of the analysis of the crystal 
structure distortions,
which are quite different in these compounds. Distortions in $ab-$plane
in LaTiO$_3$ have in general large trigonal $D_{3d}$ component. It 
gives rise to square-to-rectangle transformation in this 
plane~\cite{Cwik-03} and
results in localization of the electron on the orbital of almost 
$a_{1g}=(xy+yz+zx)/\sqrt3$ character.  
At the same time the predominant distortions in YTiO$_3$ make rhombus from
the initial square, revealing the sizable contribution of local tetragonal
distortions.

However, even in case of LaTiO$_3$ the occupied orbital
deviates from a$_{1g}$: 
\begin{eqnarray}
\label{LDAandLDAUorbitals}
 |\langle \Psi_{LDA+U}|a_{1g} \rangle|^2 = 91 \%
 \\
 |\langle \Psi_{LDA}|a_{1g} \rangle|^2 = 96  \%
\end{eqnarray}

The variation of $Ti-O$ distances in YTiO$_3$~\cite{MacLean-79} leads
to antiferroorbital ordering (Fig.~4,5)
inducing according to Goodenough-Kanamori rules the ferromagnetic structure in
agreement with experiment.\cite{Hester-97,Akimitsu-01}

The basal plane in LaTiO$_3$ undergoes the elongation in the direction
of orthorhombic $a-$axis. It leads to such kind of orbital ordering then the
orbitals of Ti ions placed in the $ab-$plane point almost along the same 
direction, but not to the oxygens (Fig.~\ref{ab-plane}, left panel).
Thus, this in-plane ``ferroorbital ordering'' 
implies large hopping integrals not only between two occupied
(AFM interactions), but also between occupied and unoccupied (FM
interactions) $d-$orbitals on different sites. 
The latter agrees with direct calculation of the exchange interaction 
parameters in Hartree-Fock approximation~\cite{Solovyev-03} 
and have to be taken into account in model calculations.

\begin{center}
\begin{figure}[h!]
 \centering
 \includegraphics[clip=true,width=0.2\textwidth]{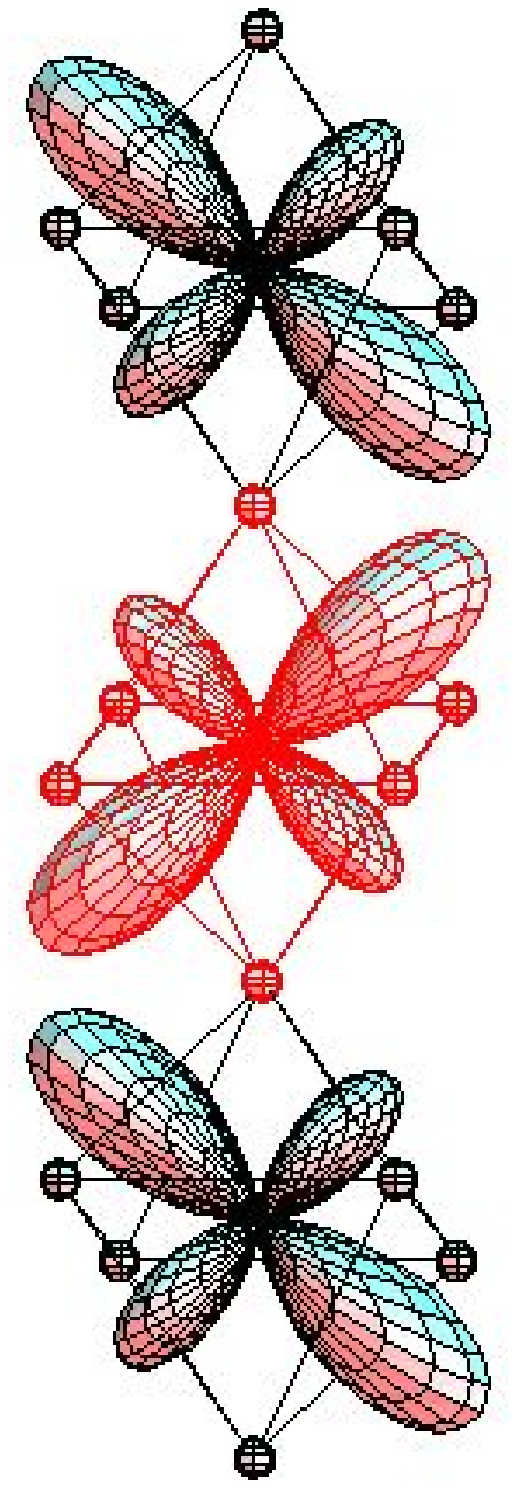}
 \includegraphics[clip=true,width=0.25\textwidth]{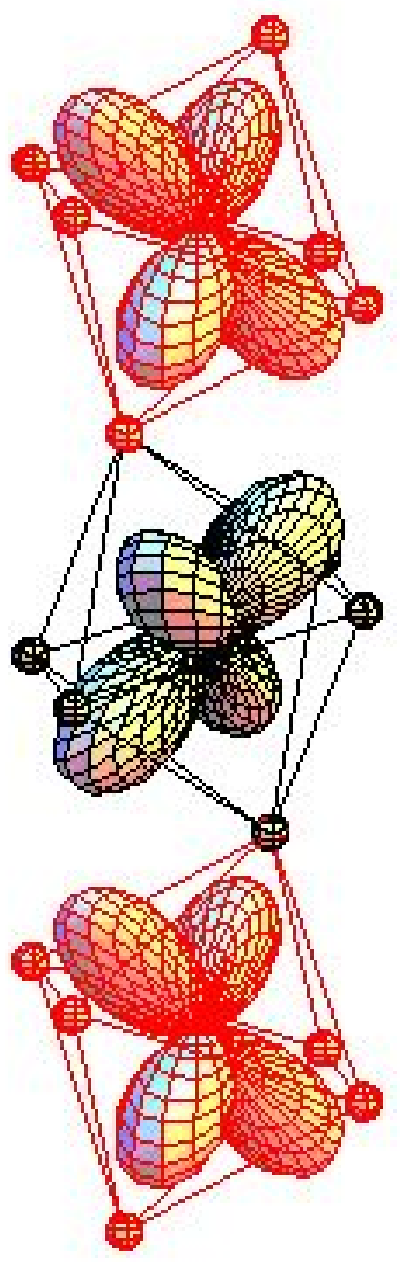}
\caption{\label{c-axis}(color online) Orbital ordering in LaTiO$_3$ (left) and 
YTiO$_3$ (right) in
$c$-direction obtained in the framework of the LDA+U calculations for AFM-G
and FM structures, respectively.}
\end{figure}
\end{center}

The presence of a mirror plane in GdFeO$_3$-type structure
perpendicular to $c-$axis~\cite{Imada-03} results in 
such kind of orbital ordering in $c-$direction that the lobes of 
the orbitals point to each other (see Fig.~\ref{c-axis}, left panel). 
This is in strong contrast to the 
orbital ordering in $ab-$plane, where orbitals are
directed away from the oxygens. 

The complicated orbital structure of LaTiO$_3$ does not favor the 
isotropic magnetic interactions in LaTiO$_3$. But this is not a unique 
situation. It is generally accepted that YTiO$_3$ shows Jahn-Teller 
distortions and corresponding orbital ordering (Ref.~\onlinecite{Sawada-97} 
and references therein). Experimentally however
it also has isotropic magnetic properties, which are hard to explain 
by orbital ordering. This would require an unrealistic set of 
parameters.\cite{Ulrich-02}

Thus the isotropic character of exchange in both LaTiO$_3$ and YTiO$_3$ 
remains an open problem. Whether the orbital 
fluctuations~\cite{Keimer-00,Ulrich-02} with 
CF splitting of $\sim$200~meV can resolve this problem, is not clear to 
us. Another option could be that the actual crystal (and consequently 
orbital) structure of LaTiO3 is somewhat different from the 
accepted one - see the recent results in Ref.~\onlinecite{Arao-02},
where the indications of monoclinic distortions have been found.
This question requires further investigations.

\vspace{-0.3cm}

\section{CONCLUSIONS}

Following the logic of the present paper we would
like to stress two general points.

(I) In the framework of LDA+U approximation we propose 
the {\it direct} method of calculation of crystal-field excitation energy. 
It can be useful for comparison with
the results of spectroscopic measurements because it takes into account
fast relaxations of electronic system. Any electron excitations in 
spectroscopy are accompanied by such relaxations and it should be 
definitely taken into consideration in the calculation of these 
excitations.

Moreover, this scheme is more reliable then others because it uses
not eigenvalues of LDA but its total energies. In addition, the method 
does not have an ambiguity in the definition of
the basis set as any projection or downfolding procedure does, because
all states included in the Hamiltonian are used. 
The latter is especially important for low symmetry systems
where it is not clear what kind of linear
combination of $d-$wave function and in which coordinate 
system should be taken for construction of a few orbital 
tight-binding Hamiltonian. 

(II) We probe this method on Y and La titanates, for which
the value of crystal-field excitation energy is indeed
essential. Using Wannier function projection of LDA full-orbital Hamiltonian 
it is found that CFS between the lowest and the next in energy $t_{2g}$ 
level is 230~meV and 180~meV for La and Y titanates, in 
agreement with previous estimate.\cite{Pavarini-04} 

According to the results of present calculations, 
fast electronic relaxations reduce CFS by $\sim$30\%
(it amounts to 70~meV in case of LaTiO$_3$) as compared with 
the difference of LDA one electron energies or CFS obtained 
by Madelung point charge summation.\cite{Cwik-03} 
However, the value of crystal-field excitation energy is still 
large enough to make orbital liquid formation in titanates rather 
unlikely. At the same time the intricate orbital ordering 
in contrast to the expectations from the model calculations
does not favor isotropic magnetic interaction in LaTiO$_3$.

\section{ACKNOWLEDGMENTS}
 We would like to thank I.V. Solovyev, D.E. Kondakov, M.A. Korotin,
A.I. Poteryaev, S. Okatov and  A.I. Lichtenstein for very useful 
discussion of calculation details. We also acknowledge fruitful 
communications with G. Khaliullin, M. Haverkort, M. Gr\"uninger, 
M. Cwik, E. M\"uller-Hartmann, M. Braden and especially L.H. Tjeng. 
This work is supported by Russian Foundation for
Basic Research under the grants RFFI-04-02-16096 and RFFI-03-0239024,
Netherlands Organization for Scientific Research through NWO 047.016.005
and by the Deutsche Forschungsgemeinschaft through SFB 608.

\end{document}